# From Naïve RAG to Deep Agentic Retrieval: An Evolving Context Engineering Pipeline for Regulatory Compliance

Mishca de Costa[*] [†], Muhammad Saleh Anwar[*] [†], Dave Mercier[*] [†], Issam Hammad [†]
[*]Data Analytics and AI, Digital Technology and Services, Ontario Power Generation, Pickering, ON, Canada
[†] The Department of Engineering Mathematics and Internetworking, Dalhousie University, Halifax, NS, Canada

***Abstract*—Retrieval-augmented generation (RAG) is the dominant paradigm for applying large language models (LLMs) to enterprise document corpora, yet naïve implementations encounter hard limits as corpus scale and query complexity grow. This paper traces the evolution of a production retrieval pipeline at Ontario Power Generation (OPG) for regulatory compliance and rate case analysis under Ontario Energy Board (OEB) reporting requirements. We examine successive stages—naïve RAG, hybrid retrieval with re-ranking, agentic function-calling retrieval, and a deep multi-agent architecture with code-based tool synthesis and explicit planning—and identify the failure modes and tradeoffs that motivated each transition. We formalize the mature architecture as Progressive Evidence Acquisition with Cost-Aware Escalation (PEA-CAE): begin with low-cost, high-precision retrieval and escalate to full-document reads only when the expected evidence gain justifies latency and cost. Our findings show that context engineering is a more tractable and economically viable path than domain-specific fine-tuning for large, evolving regulatory corpora. More broadly, the progression toward deep agentic retrieval mirrors classical information-retrieval ideas, introducing adaptive query reformulation, progressive document discovery, and hierarchical subagent summarization as practical system primitives. Operational traces further support the search-first, escalate-later pattern central to the final architecture.**

## I. Introduction

### A. Problem Domain

Large language models have demonstrated strong generalization across a wide range of language tasks, but their practical deployment in regulated industries raises immediate challenges: domain-specific corpora are large, proprietary, and continuously updated; answers must be traceable to authoritative source documents; and in jurisdictions like Canada, data residency constraints impose hard limits on which model providers and endpoints can be used for which tasks.

Ontario Power Generation faces this exact setting. As a regulated electricity generator, OPG must file detailed rate cases with the Ontario Energy Board and support stakeholders such as legal teams, engineers, and finance analysts as they navigate thousands of pages of exhibits, testimony, and supporting documentation. The question is not merely *can an LLM answer questions about this corpus* but *how do we engineer a system that reliably, efficiently, and verifiably retrieves the right context to enable correct answers*?

### B. Why Context Engineering Over Fine-Tuning

While domain-specific fine-tuning is attractive in principle, it is often economically and operationally mismatched to large evolving regulatory corpora. The models required for strong language understanding are expensive to adapt at enterprise scale, and fine-tuned parameters freeze knowledge that continues to change through new filings, amendments, and supporting evidence. This motivates a context engineering approach in which general-purpose models are supplied with carefully constructed inference-time context through retrieval, filtering, and compression.

### C. A Recurring Failure Mode: Treating Context as a Free Resource

A central empirical finding running through all stages of this work is that context length is not a free resource. Larger context windows do not imply proportional recall of relevant facts. As context grows, irrelevant tokens compete for attention, retrieval signal weakens, and answer quality can degrade despite higher cost and latency. We define **context rot** as the quality loss from overextended, noisy context: key facts remain present but become increasingly difficult for the model to retrieve and prioritize during generation.

This observation motivates a formal objective that runs throughout the paper: maximize relevant evidence density in the context window, not raw context volume.

### D. Contributions

This paper makes the following contributions, summarized schematically in Fig. 1:

1) A stage-by-stage algorithmic characterization of three retrieval architectures deployed in a production regulatory AI system, connected to the failure modes that drove each transition.
2) A formal framework, **PEA-CAE** (Progressive Evidence Acquisition with Cost-Aware Escalation), capturing the retrieval policy of the mature architecture.
3) An empirical analysis of context window utilization and retrieval precision tradeoffs at each stage, including operational trace data.

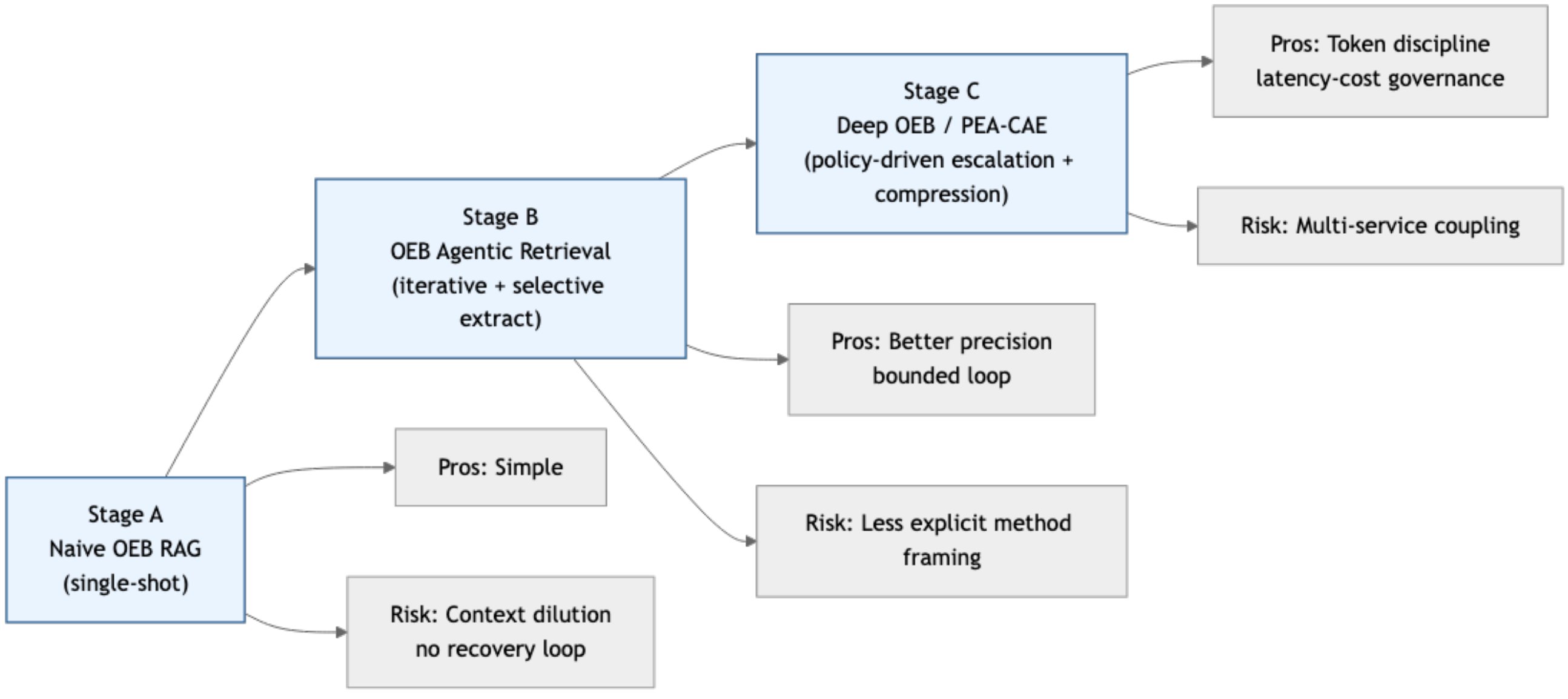

Fig. 1. Method evolution across stages, from naïve RAG to deep agentic retrieval.

4) A treatment of data sovereignty constraints as a first-class architectural consideration, yielding a capability partitioning pattern for multi-jurisdiction deployments.
5) A comparative analysis connecting practical system design decisions to classical information retrieval theory.

## II. Background and Related Work

### A. Retrieval-Augmented Generation

Retrieval-augmented generation (RAG) combines retrieval over an external corpus with answer generation conditioned on the retrieved evidence [1]. Naïve RAG treats retrieval as a fixed, stateless lookup in which top-$k$ chunks are retrieved once and passed directly to the model, an approach that becomes limiting as corpus size and query complexity increase. Recent work in enterprise contexts highlights that standard chunk boundaries can severely limit synthesizer performance, prompting the development of atomic-unit and question-based retrieval paradigms [2]. In highly specialized regulatory domains, RAG-powered architectures have been shown to significantly outperform direct long-context prompting when analyzing lengthy compliance documents [3]. Similarly, within the nuclear domain, Anwar *et al.* report improvements from retrieval-augmented prompting over prompt-only baselines on domain-specific data [4].

### B. Context Engineering and the Long-Context Recall Problem

Context engineering refers to the deliberate construction of high-signal context for LLM inference through retrieval, filtering, and compression. A key empirical constraint is that effective recall does not scale linearly with context window size: as irrelevant tokens accumulate, attention competition can degrade answer quality even in long-context models [5]. To combat this phenomenon, comprehensive prompt compression frameworks [6] and tools such as LLMLingua [7] have been introduced to algorithmically prune irrelevant tokens and preserve reasoning signals. We formally encapsulate this optimization objective through the metric of context efficiency:

$$\eta = \frac{R}{B} \tag{1}$$

where $R$ is the token count of genuinely relevant content and $B$ is the total context budget consumed.

### C. Agentic Retrieval and Tool Use

Tool-using language models can improve retrieval by iteratively reformulating queries, selecting among tools, and reasoning over intermediate results [8], [9]. Frameworks such as Self-RAG [10] formalize this via self-reflective tokens that allow the model to critique and refine retrieved passages on-demand. This connects naturally to classical adaptive information retrieval [11] and information foraging theory [12], where search proceeds through progressive discovery rather than static lookup. As enterprise workflows scale, these concepts evolve into fully Agentic RAG architectures [13], where multiple specialized agents collaborate on planning, tool use, and multi-step reasoning. Function-calling LLMs have also been successfully applied to nuclear plant data retrieval through natural language-to-SQL tool invocation [14].

### D. Document Processing for Regulatory Corpora

Regulatory filings are predominantly distributed as unstructured PDFs, making extraction fidelity a foundational dependency for downstream retrieval. OCR accuracy, table parsing, and layout handling directly set the quality ceiling for indexed text. Because regulatory exhibits frequently rely on charts, schedules, and infographics, traditional text-only extraction is often insufficient; this necessitates multimodal RAG approaches that align visual and textual data to ground generative outputs [15]. Related work on LLM-based classification of nuclear safety events further illustrates the broader applicability of LLMs to operational text tasks in regulated energy settings [16-17].

## III. Problem Setting and Corpus Characteristics

### A. Task Formulation

Given a user query $q$, produce an answer $a$ that is:

- grounded in admissible OEB evidence,
- scoped correctly to the relevant filing structure (submission, exhibit, rate case year),
- efficient enough for interactive or near-interactive use.

### B. Corpus Structure

The OEB corpus is organized by rate case year and exhibit number, and includes testimony, financial schedules, engineering studies, and interrogatory responses. Its main design-relevant properties are scale, heterogeneous content types, semantically meaningful document hierarchy, and temporal versioning, all of which affect scoped retrieval and answer correctness.

### C. Query Complexity Distribution

Queries range from simple factual lookups to multi-hop comparative questions spanning exhibits and rate case years, motivating retrieval strategies that can adapt in depth and scope.

### D. System Constraints

The system operates under four constraints that function as first-class architectural parameters:

1) **Precision under governance scope:** Retrieval must respect filing structure; unconstrained cross-case search can produce legally incoherent evidence mixtures.
2) **Escalation discipline:** Full-document extraction is expensive and should be triggered selectively based on expected evidence gain.
3) **Loop boundedness:** Multi-step reasoning must terminate under explicit budget or confidence criteria.
4) **Data residency:** Operating in Canada, OPG cannot permit sensitive regulatory data to be transmitted to cloud services hosted outside Canada. At the time of deployment, certain LLM provider capabilities—specifically multimodal vision inference—were available only through non-Canadian endpoints, making this constraint architecturally consequential.

## IV. Stage I— Naïve RAG Baseline

### A. Ingestion Pipeline

The initial pipeline follows a standard RAG ingestion flow. Documents are converted to text through OCR or structured extraction, segmented into 500–800 token chunks, annotated with metadata such as exhibit and rate case year, embedded, and stored in a vector index. The chunk size was chosen to balance semantic coherence, embedding limits, and retrieval granularity.

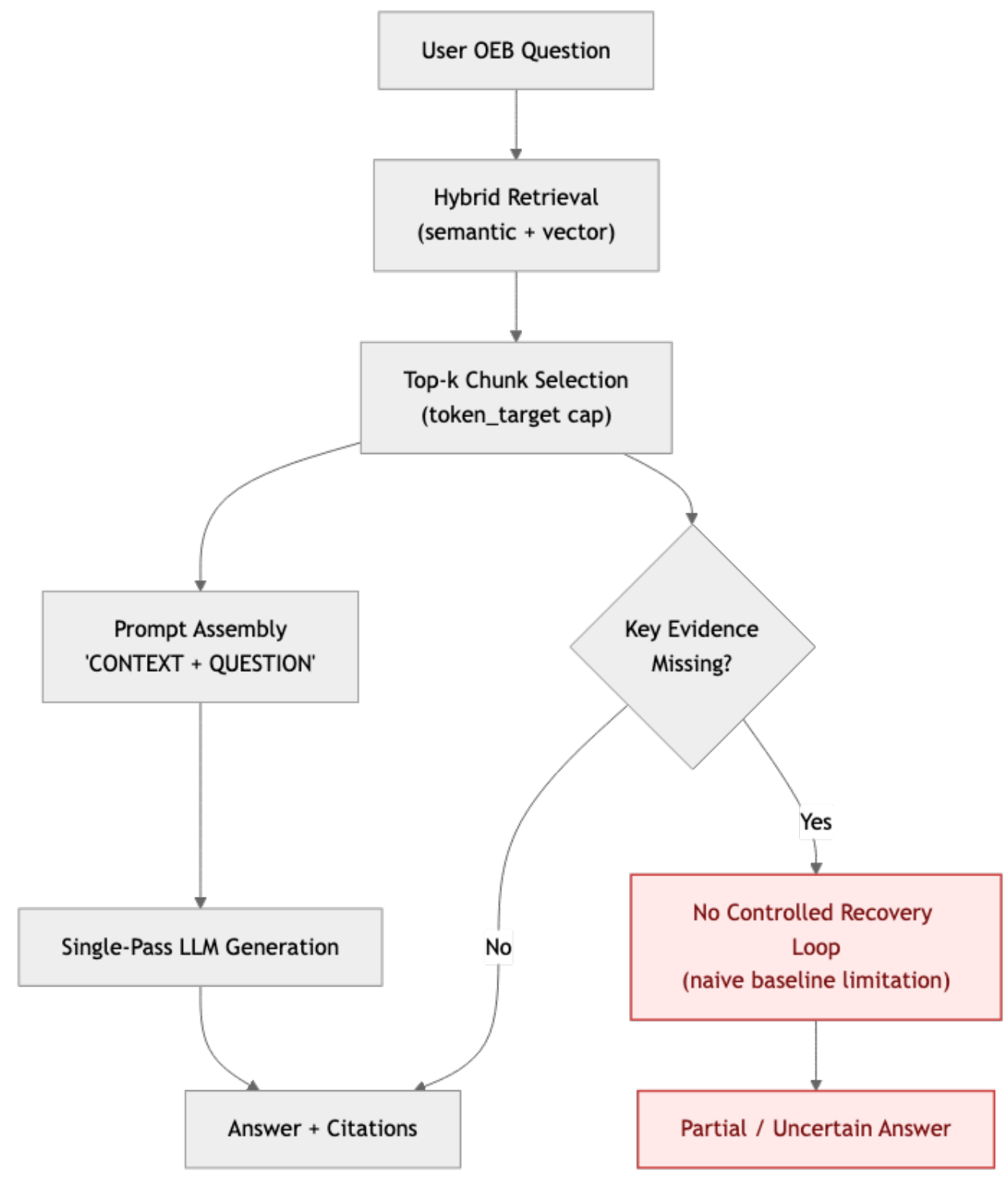

Fig. 2. Stage I architecture: Naïve RAG with hybrid retrieval and re-ranking.

### B. Retrieval and Context Assembly

At query time, the system performs hybrid retrieval combining dense vector similarity with keyword matching, followed by cross-encoder re-ranking. The top-ranked results, typically yielding 15,000–25,000 tokens of context, are assembled with the query and system prompt for answer generation. This stage is deterministic: identical queries produce identical retrieved contexts, which simplifies debugging but prevents the system from adapting when retrieval fails. The overall Stage I retrieval pipeline is illustrated in Fig. 2.

### C. Context Window Evolution

Larger context windows improved recall relative to earlier deployments, but effective retrieval remained bounded well below nominal frontier context limits. In practice, selective retrieval remained necessary because increasing context volume also increased cost, latency, and attention competition.

### D. Failure Modes

The na¨ıve RAG architecture has a fundamental rigidity: the retrieval step is deterministic and cannot self-correct. When the re-ranker fails to surface the most relevant chunks—due to sub-optimal chunk boundaries, vocabulary mismatch, or multi-hop information dependencies—the LLM has no recourse. The only mitigation is to increase $k$ (more retrieved chunks), which directly compounds cost and latency. Formally, this means the system cannot adaptively improve context efficiency $\eta$;

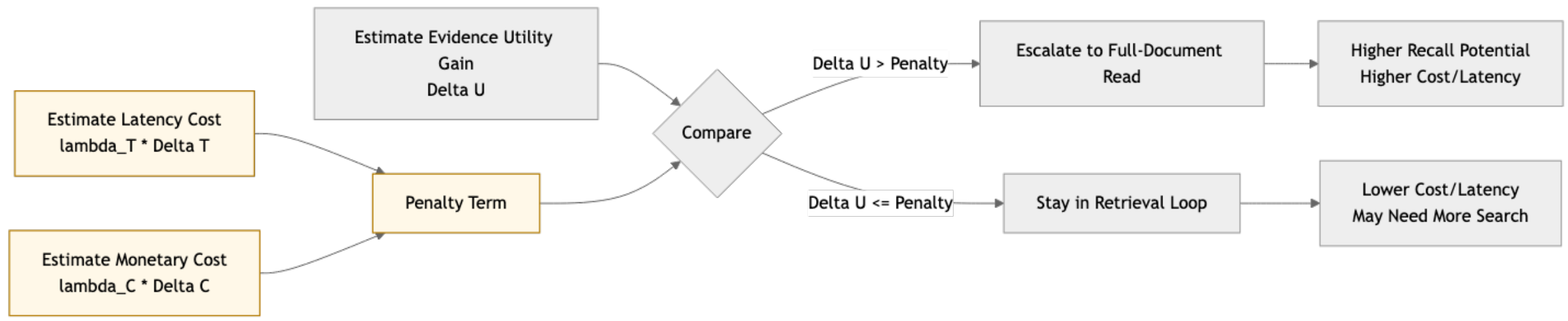

Fig. 3. Escalation decision rule: move from chunk retrieval to full-document reading when expected evidence gain outweighs latency and cost.

it can only buy more $B$ in hopes of capturing more $R$, with diminishing returns.

Beyond recall quality, large context windows introduce concrete costs: token processing costs scale linearly with context length, inference latency increases substantially, and at the upper end, context payloads approach limits of HTTP request body sizes—an early systems-level signal that the "stuff more tokens" strategy had structural limits.

## V. Stage II—Agentic Retrieval via Function Calling

### A. Motivation: From Passive Consumer to Active Retriever

The core insight motivating this stage is that a capable LLM should not be a passive consumer of a fixed retrieved context—it should be an active participant in the retrieval process. Given a tool that exposes the search index, an LLM agent can iteratively probe the corpus, refine its queries based on intermediate results, and converge on the information it needs through **progressive discovery**. This transforms retrieval from a deterministic lookup into a heuristic search process.

### B. Progressive Discovery Algorithm

The underlying hybrid + re-ranked search index from Stage I is retained. What changes is the control flow:

- **Tool Exposure:** The LLM is provided a retrieval tool: a callable function accepting a query string, optional metadata filters (exhibit ID, rate case year), and a top-$k$ parameter, returning matching chunks.
- **Iterative Query Reformulation:** Rather than submitting the user's raw query once, the LLM issues multiple tool calls with reformulated queries. Each call returns a small number of chunks—typically $k = 5$—which the LLM evaluates before deciding whether to refine, expand, or pivot its search strategy. This mirrors pseudo-relevance feedback in classical IR, but with the LLM driving the feedback loop rather than a statistical model.
- **Filter-Guided Search:** The LLM is informed in its system prompt of the corpus structure: which exhibits exist, what each rate case contains, which document types tend to hold which kinds of information. Armed with this knowledge, the LLM scopes queries to specific exhibits or years, reducing the search space and improving precision. This structured metadata functions as **semantic routing** over the corpus. Critically, this tuning is not merely a usability concern—it is a direct token efficiency optimization. A well-tuned filter-aware agent can reach the answer with fewer total tokens than the fixed $k = 25{,}000$ approach, because it retrieves only informative chunks.

### C. Chunk-to-Document Escalation Policy

When chunk-level search returns signals pointing strongly to a particular source document, the agent can retrieve the **full document** rather than continuing to probe with additional chunk searches. A subordinate agent—or a smaller, cost-efficient model—reads the full document and returns a targeted summary or specific passages to the main agent. This adaptive granularity (moving from chunks to whole documents when the evidence warrants) mirrors the patch exploitation behavior in information foraging theory: once a high-value region of the information landscape is identified, the agent switches from broad scanning to deep exploitation of that patch. The escalation logic is illustrated in Fig. 3.

This two-tier policy can be expressed as a decision rule: escalate to full-document retrieval if and only if the expected evidence gain $\Delta U$ justifies the added latency $\Delta T$ and cost $\Delta C$:

$$\text{Escalate} \iff \Delta U > \lambda_T \Delta T + \lambda_C \Delta C \tag{2}$$

where $\lambda_T$ and $\lambda_C$ are penalty weights. At $k = 5$ chunks, the agent quickly identifies high-confidence document targets; the escalation decision then replaces many further chunk searches with a single document read—often a net efficiency gain.

This adaptive granularity—moving from chunks to whole documents when the evidence warrants—mirrors the patch exploitation behavior in information foraging theory: once a high-value region of the information landscape is identified, the agent switches from broad scanning to deep exploitation of that patch.

### D. Capability Partitioning for Data Residency

Because full-document retrieval exposes charts, graphs, and figures requiring vision understanding, and because the primary LLM provider's vision endpoint does not satisfy Canadian data residency requirements, the pipeline routes document-level visual content to a vision-capable model hosted within Canada. This model is less capable as a general reasoner but sufficient for extracting structured information from visual elements; its outputs are returned to the main agent as text.

This **capability partitioning** pattern—selecting models by functional fitness within jurisdictional constraints rather than purely by capability—generalizes to any multi-jurisdiction deployment in healthcare, finance, or government contexts.

### *E. Token Overhead and the "Too Many Tools" Problem*

Function-calling introduces per-call overhead in the form of JSON schema tokens for tool definitions, call serialization, and result formatting. In the worst case—many low-quality queries failing to converge—total token consumption can exceed the Stage I baseline. This motivates the transition to a deeper agent architecture in Stage III, where tool overhead is amortized through code-based tool synthesis. The overall Stage II architecture is summarized in Fig. 4.

## VI. Stage III—Deep Agent Retrieval with Planning and Code Tools

### *A. Explicit Planning and Task Decomposition*

The key capability introduced in Stage III is explicit planning: before retrieval begins, the agent decomposes the user query into a sequence of sub-questions, orders them by expected evidence utility, and stores this as an explicit data structure within its working memory, checking off completed sub-tasks and recording intermediate findings.

This planning step transforms retrieval from a reactive process (answer the question as posed) into a deliberate research process (understand what information is needed, locate it systematically, synthesize it). For multi-hop queries—comparing cost justifications across two rate case years, cross-referencing capital program schedules with revenue calculations—structured decomposition is essential for completeness and auditability. The overall Stage III architecture is illustrated in Fig. 5.

### *B. Hierarchical Subagent Architecture and Context Rot Prevention*

A single LLM agent is bounded by its context window. Extended research tasks that require collecting and synthesizing large amounts of information will eventually exhaust available context, and the accumulation of intermediate tool results degrades attention quality over long reasoning chains.

The deep agent architecture addresses this through **subagent delegation**: the main agent spawns subordinate agents to handle bounded sub-tasks (e.g., "read exhibit C-2 and extract all capital expenditure line items for 2022–2024"). The sub-agent conducts its research—potentially issuing many retrieval calls and accumulating significant intermediate context—and returns only its **synthesized output** to the main agent. The tokens burned by the subagent during research are not added back into the main agent's conversation history. This implements **hierarchical summarization at the agent level**: information is progressively compressed as it moves up the delegation hierarchy, preserving high-level findings while discarding intermediate retrieval artifacts. Formally, let the main agent's context at step $t$ be $C_t$ and let a subagent burn $\delta$ tokens during its research. The subagent returns a compressed summary $s$ of size $|s| \ll \delta$; only $s$ is added to $C_t$. This allows the system to sustain research tasks of 10–20 minutes in duration—far beyond what a single LLM call or single-agent loop could achieve within context constraints.

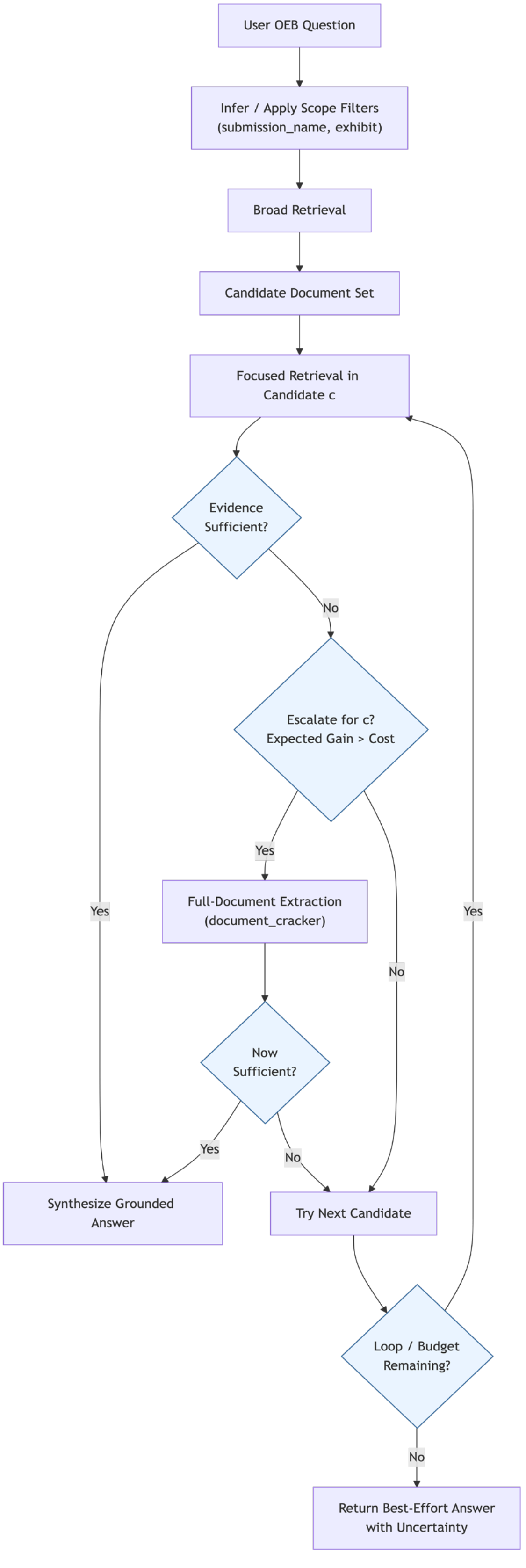

Fig. 4. Stage II architecture: agentic retrieval via function-calling and progressive discovery.

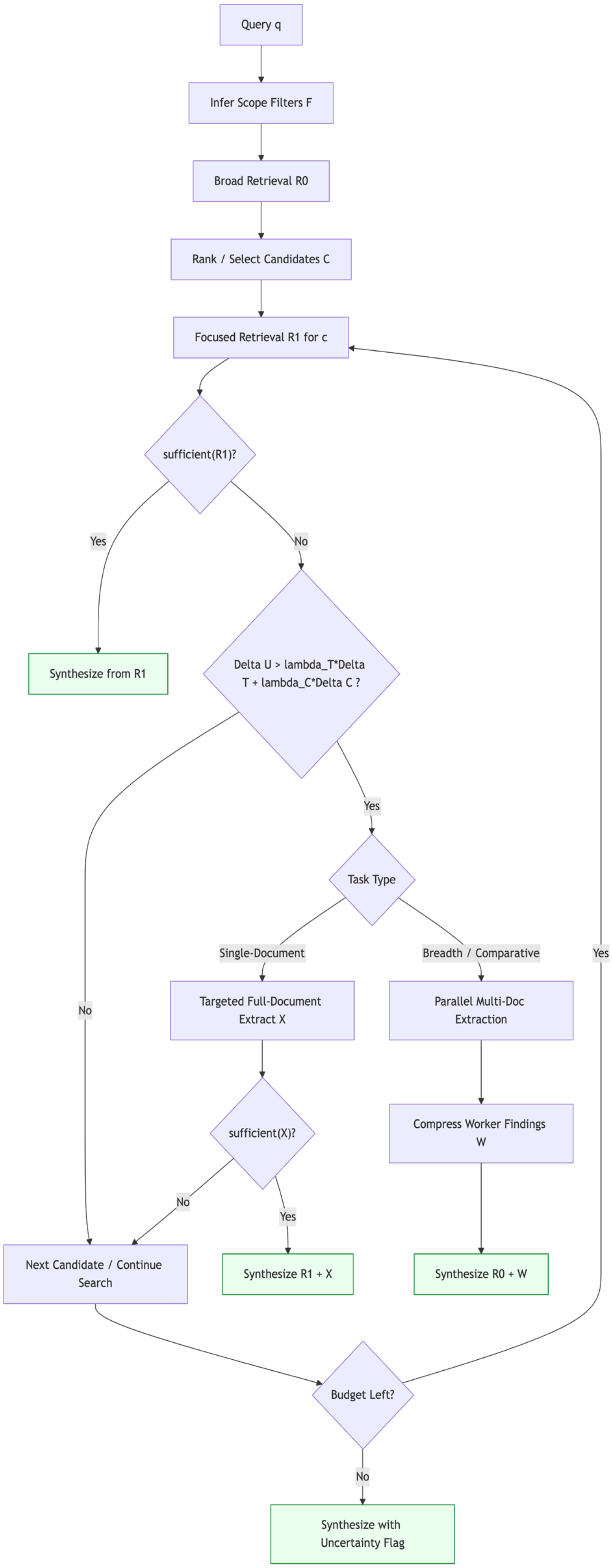

Fig. 5. Stage III architecture: deep agentic retrieval with planning, subagents, and code-based tools (PEA-CAE).

The main agent also employs **automatic summarization** of earlier conversation segments as the context window fills, trading some information fidelity for extended operational duration.

### C. Code-Based Tool Synthesis

Instead of pre-defined JSON function tools, the deep agent is given access to a **code execution environment** (Python and Bash). The agent writes code at runtime to interact with search APIs, process retrieved data, perform calculations, and orchestrate sub-tasks. This approach has several algorithmic advantages:

- **Combinatorial Expressiveness:** A single code environment subsumes an unbounded number of narrow tools. Rather than pre-engineering every possible retrieval operation as a named function, the agent composes operations programmatically using API documentation provided in its context.
- **Reduced Schema Overhead:** Code execution does not require JSON tool-call serialization for every individual operation. The overhead is amortized across the program rather than paid per-call, directly addressing the JSON overhead problem identified in Stage II.
- **Precision:** Programmatic logic enables more precise retrieval and post-processing than natural-language tool invocations. Complex operations can run in a single code block rather than multiple sequential calls.

The tradeoff is that code execution requires a sandboxed runtime and that the LLM must produce syntactically and semantically correct code. This is viable with frontier models for the retrieval and data manipulation tasks encountered in this domain, but represents a capability threshold that limits applicability to less capable models.

### D. Extended Task Horizon

By splitting work across multiple bounded LLM calls—each ending in a summarization or handoff step rather than a hard context limit—the deep agent can sustain research tasks that would be impossible for a single inference call. In the OEB deployment, typical complex research tasks run for 10–20 minutes, with some queries requiring up to an hour. This temporal scaling, achieved through planning, subagent delegation, and summarization, is a direct consequence of the context rot prevention mechanisms described above.

## VII. Formal Framework: PEA-CAE

Drawing together the algorithmic primitives from Stage III, we formalize the mature retrieval architecture as **Progressive Evidence Acquisition with Cost-Aware Escalation (PEA-CAE)**.

### A. Retrieval Score

For document chunk $d$ and query $q$, the hybrid retrieval score is:

$$S(d \mid q) = \alpha \cdot S_{\text{semantic}}(d, q) + (1 - \alpha) \cdot S_{\text{keyword}}(d, q) \quad (3)$$

with an admissibility gate $f(d) = 1$ enforcing OEB scope constraints (exhibit, rate case year). Only admissible chunks enter the candidate set.

### B. Escalation Decision Rule

$$\text{Escalate to full document} \iff \Delta U > \lambda_T \cdot \Delta T + \lambda_C \cdot \Delta C \quad (4)$$

where $\Delta U$ is the expected evidence utility gain from full-document reading, $\Delta T$ is added latency, $\Delta C$ is added cost, and $\lambda_T$, $\lambda_C$ are penalty weights calibrated to operational requirements.

### C. Context Engineering Rules

The PEA-CAE framework applies four context management rules as first-class algorithmic constraints:

1) **Filter first:** constrain the search space using structured metadata before adding evidence to context.
2) **Compact early:** represent intermediate search evidence in concise structured form to preserve context budget for reasoning.
3) **Escalate late:** read full documents only when chunk evidence fails an explicit sufficiency criterion.
4) **Compress breadth:** in multi-document analysis tasks, aggregate subagent findings before final synthesis to prevent context growth proportional to document count.

### D. Algorithm Sketch

Algorithm 1 summarizes the PEA-CAE retrieval policy as a staged procedure that first constrains scope, then performs low-cost retrieval, and escalates to full-document extraction only when the expected evidence gain justifies added latency and cost. The final branch handles breadth-oriented tasks through parallel extraction and compression before synthesis.

**Algorithm 1** PEA-CAE

1: **Require:** query $q$
2: Plan $\leftarrow$ construct_research_plan($q$)
3: $F \leftarrow$ infer_scope_filters($q$) ▷ exhibit ID, rate case year
4: $R_0 \leftarrow$ broad_retrieval($q, F, k = 5$)
5: $C \leftarrow$ candidate_documents($R_0$)
6: **for all** $c \in C$ **do**
7: $R_1 \leftarrow$ focused_retrieval($q, c, F, k = 5$)
8: **if** sufficient($R_1$) **then**
9: **return** synthesize($R_1$)
10: **if** $\Delta U(c, q) > \lambda_T \, \Delta T(c) + \lambda_C \, \Delta C(c)$ **then**
11: $X \leftarrow$ full_document_extract($q, c$) ▷ via subagent or vision model
12: **if** sufficient($X$) **then**
13: **return** synthesize($R_1, X$)
14: **if** breadth_task($q$) **then**
15: $W \leftarrow$ parallel_extract_and_compress($q$, top_docs($C$))
16: **return** synthesize($R_0, W$)
17: **return** synthesize($R_0$, uncertainty = true)

## VIII. Empirical Observations

Operationally, total latency and cost are driven by the number of retrieval iterations, the frequency of full-document escalation, and the degree of parallel subagent execution. The following observations are drawn from operational traces of the Stage III (deep agent) system. Stage I (naïve RAG) behavior is characterized methodologically from deployment experience; direct comparative benchmarks across all stages under identical query sets are reserved for future work.

### A. Run Outcomes

From a sample of 14 system runs across query types:

- OEB agentic runs: 6 of 6 successful
- Observed task durations: min 24.8 s, median 92.1 s, max 223 s

These durations reflect the full research loop including planning, iterative retrieval, and synthesis, consistent with the design intent of a sustained research pipeline rather than a single-pass lookup.

### B. Retrieval-to-Escalation Ratio

From OEB operational logs:

- Chunk retrieval calls: 67
- Full-document extraction calls: 7
- Search-to-escalation ratio: ≈9.6:1

This ratio is consistent with the PEA-CAE design intent: escalation to full-document reads is selective, occurring in roughly 10% of retrieval events. The majority of queries are resolved at the chunk level, with escalation reserved for cases where the chunk-level evidence is insufficient.

### C. Extraction Tail Behavior

Observed full-document extraction latencies ($n = 7$):

- Min: 0.70 s, Median: 2.44 s, Max: 12.14 s

The tail behavior (≈ 12× spread from min to max) confirms why escalation must remain selective: at the tail, a single document extraction consumes as much time as several chunk retrieval passes.

## IX. Comparative Analysis

Table I summarizes the progression across the three retrieval stages. The comparison makes clear that retrieval flexibility and support for multi-hop questions improve across stages, but at the cost of greater orchestration complexity and less predictable latency.

TABLE I
COMPARISON OF RETRIEVAL STAGES

| Dimension | Stage I: Naïve RAG | Stage II: Agentic (Function Call) | Stage III: Deep Agent (PEA-CAE) |
|---|---|---|---|
| Retrieval Control | Fixed, deterministic | Iterative, LLM-directed | Planned, policy-driven, multi-agent |
| Query Strategy | Single pass, raw user query | Iterative reformulation, $k = 5$ probing | Decomposed plan, prioritized sub-queries |
| Context per Query | 15K–25K tokens (fixed) | Adaptive (often fewer for simple queries) | Adaptive + subagent compression |
| Scope Control | Static index-level filters | Filing filters available in loop | Filters as primary governance gate |
| Multi-hop Queries | Poor (no recovery path) | Moderate | Strong |
| Visual Content | Absent (data residency) | Offloaded to CA-resident model | Offloaded to CA-resident model |
| Context Rot Risk | Low (single pass) | Moderate (accumulates over tool calls) | Low (subagent compression) |
| Tool Scalability | N/A | Bounded by tool recall capacity | Unbounded via code synthesis |
| Task Horizon | Single inference call | Minutes | 10–20+ minutes |
| Latency | Low (single retrieval pass) | Variable | Higher for complex queries |
| Token Cost | Predictable | Variable (lower or higher than Stage I) | Higher on complex, lower on simple |
| Engineering Complexity | Low | Moderate | High |

## X. DISCUSSION

The progression from naïve RAG to deep agentic retrieval suggests that, for large regulated document corpora, system performance is determined less by domain-specific model adaptation than by retrieval policy and context construction. Across all stages, the dominant pattern is consistent: better outcomes come not from maximizing context volume, but from maximizing relevant evidence density through scoped retrieval, selective escalation, and aggressive compression. This reinforces the view that context engineering is not merely prompt design, but a broader algorithmic discipline concerned with how evidence is discovered, filtered, structured, and delivered to the model at inference time.

The architecture is also shaped by governance constraints, particularly data residency, which in this setting motivates capability partitioning across model endpoints rather than a single-model solution, aligning with broader efforts to deploy privacy-preserving models in the nuclear sector [18]. More broadly, the transition from fixed retrieval to adaptive, planned, multi-agent retrieval reflects a practical reality of enterprise AI systems: retrieval quality, tool overhead, latency, and context growth must be managed jointly rather than optimized in isolation. The mature system therefore embodies a simple but durable principle: retrieve progressively, escalate selectively, and compress aggressively, because context remains a scarce resource even when nominal context windows continue to grow.

## XI. CONCLUSION

This paper has presented a stage-by-stage account of the evolution of a production RAG pipeline at Ontario Power Generation, from naïve hybrid search to a deep agentic retrieval system with code-based tool synthesis, explicit re-search planning, and hierarchical subagent delegation. Each stage was motivated by concrete failure modes in the prior architecture, and each introduced algorithmic innovations—adaptive query reformulation, filter-guided semantic routing, chunk-to-document escalation, capability partitioning for data residency, and subagent-mediated context compression—that address those failures.

The central finding is that context engineering, rather than model fine-tuning, is the right paradigm for large, evolving, regulated-industry document corpora. The intelligence of such a system lies not in the model weights but in the algorithms that construct high-signal context windows—and the evolution of those algorithms mirrors the progression from passive retrieval to active, planned, cost-aware information foraging. The PEA-CAE formalization captures the mature architecture's core insight: retrieve progressively, escalate selectively, and compress aggressively—because context is not free. A con-trolled comparative benchmark across retrieval stages remains an important direction for subsequent work.

## XII. ACKNOWLEDGMENT

This research paper was supported by Ontario Power Gen eration (OPG) and by The Natural Sciences and Engineering Research Council of Canada (NSERC) and The Canadian Nuclear Safety Commission (CNSC) grant number ALLRP-580442-2022.